\documentstyle[12pt]{article}
\newcommand{\be}{\begin{equation}}
\newcommand{\ee}{\end{equation}}
\newcommand{\ba}{\begin{eqnarray}}
\newcommand{\ea}{\end{eqnarray}}
\def\n{\noindent}

\catcode `\@=11
\@addtoreset{equation}{section}
 
\catcode `\@=12

\title{\bf\huge Higher Dimensional Analogue of McVittie Solution                                     }     
\author{L.K. Patel \\
{\sl Department of Mathematics,}\\
{\sl Gujarat University, Ahmedabad 380 009. India} \\
R. Tikekar \\
{\sl Department of Mathematics,}\\
{\sl Sardar Patel University, Vallabh Vidyanagar 388 120, India.} \\
Naresh Dadhich\thanks{E-mail : nkd@iucaa.ernet.in} \\
{\sl Inter-University Centre for Astronomy \& Astrophysics,}\\
{\sl Post Bag 4, Ganeshkhind, Pune - 411 007, India.}
}
\date{}
\begin{document}
\maketitle

\begin{abstract}
A generalization of the McVittie solution, representing
spacetime of a mass particle placed in $(n+2)$ dimensional
Robertson-Walker universe is reported.
\end{abstract}
\newpage
\n McVittiee [1] obtained a solution of the Einstein's field equations
which is described by the metric

\be
ds^2 = \bigg[ \frac{1 - \mu(r,t)}{1 + \mu(r,t)} \bigg]^2 dt^2
- \frac{[1+ \mu(r,t)]^4}{(1+k \frac{r^2}{4})^2} R^2(t) (dr^2
+ r^2 d \theta^2 + r^2 sin^2 \theta^2 + r^2 sin^2 \theta d \phi^2)
\ee

\n where

\be
\mu(r,t) = \frac{m}{2rR(t)} (1+k \frac{r^2}{4})^{1/2},
\ee

\n $m$ is a constant and $k=0, \pm 1$. This solution which is important 
for historical reasons is a superposition of the Schwarzschild exterior
solution and the Robertson-Walker solution. It represents the gravitational
field of mass particle placed in a Robertson Walker Universe. Several
aspects of the McVittiee solution, of geometrical and physical relevance
have been investigated by many authors [2]. Raychaudhuri [3] established
the uniqueness of McVittiee solution under following assumptions:

(i)	The metric is spherically symmetric with a singularity at the centre.

(ii)	The matter distribution is a perfect fluid.

(iii)	The metric must asymptotically go over to the isotropic 
cosmological form.

(iv)	The fluid flow is shear-free.

Developments in superstring theories [4-5] have stimulated the study
of physics in higher dimensional spacetimes [6]. 
Subsequently the implications of relativistic field equations
in which dimensions of the spacetime are $1+D, 3 \leq D \leq 9$, in spite of 
the lack of direct observational evidence in favour of extra dimensions
so far, have been studied by several authors [7-9]. Solutions
of Einstein field equations in higher dimensional spacetimes are
believed to be of physical relevance possibly at the extremely
early times before the universe underwent the compactification
transitions. Liddle et al [10] have discussed the consequences
of  extra dimensions on the structure and maximum mass of a
star.. However to our knowledge non-static models describing spacetime
of a star placed in a higher dimensional, homogeneous and isotropic universe
have not been studied so far. Therefore we thought it worthwhile to 
obtain a higher dimensional counterpart of the McVittiee solution.

We begin with the non-static spherically symmetric metric

\be
ds^2 = e^{\nu(r,t)} dt^2 - e^{w(r,t)} \lbrace dt^2 + r^2 d \Omega^2_n \rbrace
\ee

\n where

\be
d \Omega^2_n = d \theta^2_1 + sin^2 \theta_1 d \theta^2_2 + \ldots
+ sin^2 \theta_1 sin^2 \theta_2 \ldots sin^2 \theta_{n-1} d \theta^2_n
\ee

\noindent of a $(n+2)$ - dimensional spacetime in isotropic coordinates and

\be
T^i_k = (\rho + p) u^i u_k - p \delta^i_k.
\ee

\noindent the energy-momentum tensor for perfect fluid describing its 
physical content. Adopting co-moving coordinates so that

\be
u_i = e^{\nu/2} \delta^t_i,
\ee

\n   the system of Einstein's field equations reduces to the following
set of equations :

\be
8 \pi \rho = \frac{n(n+1)}{8} e^{- \nu} \dot w^2 - \frac{n}{2}
e^{-w} [ w^{\prime \prime} - \frac{(n-1)}{4} w^{\prime^2}
+ \frac{n}{2} \frac{w^{\prime}}{r} ]
\ee

\be
8 \pi p = \frac{n}{2} e^{-w} [\nu^{\prime} ( \frac{w^{\prime}}{2} +
\frac{1}{r}) + \frac{(n-1) w^{\prime}}{r} + \frac{(n-1)}{4} w^{\prime^2}
] - \frac{n}{2} e^{- \nu} [ \ddot w - \frac{\dot w \dot \nu}{2} +
\frac{(n+1)}{4} \dot w^2 ]
\ee

\be
\dot w^{\prime} - \frac{\dot w \nu^{\prime}}{2} = 0
\ee

and

\be
\nu^{\prime \prime} + (n-1) w^{\prime \prime} + \frac{\nu^{\prime 2}}{2}
- \frac{(n-1)}{2} w^{\prime 2} - w^{\prime} \nu^{\prime} -
\frac{\nu^{\prime}}{r} - (n-1) \frac{w^{\prime}}{r} = 0.
\ee
 
\n Here and in what follows an overhead prime and an overhead dot respectively
denote differentiations with respect to $r$ and $t$.

Higher dimensional analogue of the Robertson-Walker metric follows on setting
\be
\nu = 0, ~e^w = \frac{R^2(t)}{(1+k \frac{r^2}{4})^2}.
\ee
\n The higher dimensional analogue of the exterior schwarzschild metric is 
in isotropic coordinates is obtained on setting
\be
e^{\nu} = \bigg( \frac{1 - \frac{m}{2(n-1) r^{n-1}}}{1 + \frac{m}{2(n-1) 
r^{n-1}}} \bigg)^{2}, ~e^w = \bigg(1 + \frac{m}{2(n-1) r^{n-1}}  
\bigg)^{4/(n-1)}. 
\ee

\n The expressions (1), (11) and (12) suggest that the metric (3) with
\be
e^{\nu} = \bigg\lbrace \frac{1-B}{1+B} \bigg\rbrace^2, ~e^w = 
(1+B)^{4/(n-1)} \frac{R^2}{(1+k \frac{r^2}{4})^2}
\ee
\n may lead to the higher dimensional version of McVittie's solution. 
Here $B$ is a function of $r$ and $t$. In view of (13) equations (9) and 
(10) become
\be
(1-B) \frac{\dot B^{\prime}}{B^{\prime}} + \dot B + (n-1)
\frac{\dot R}{R} = 0
\ee
\n and
\be
-\frac{B^{\prime \prime}}{B} + \frac{n+1}{n-1}. \frac{B^{\prime 2}}
{B^2} + (\frac{1-\frac{3}{4} kr^2}{1+\frac{k}{4} r^2}) \frac{B^{\prime}}{rB}
= 0.
\ee
\n The equations (14) and (15) admit the solution

\be
B = \frac{m}{2(n-1)} r^{1-n} R^{1-n} (1+ \frac{k}{4} r^2)^{(n-1)/2}
\ee
\n where $m$ is an arbitrary constant. Subsequently the metric
\be
ds^2 = (\frac{1-B}{1+B})^2 dt^2 - (1+B)^{\frac{4}{n-1}}
\frac{R^2(t)}{(1+ k \frac{r^2}{4})^2} (dr^2 + r^2 d \Omega^2_n)
\ee
\n with $B$ as given by (16) represents higher dimensional analogue
of the McVittiee solution. The pressure and density of the fluid
filling this universe are

\ba
\frac{16 \pi p}{n} &=& - \frac{2(1+B)}{(1-B)} . \frac{\ddot R}{R}
- \frac{(n-1) - (n+3) B}{1-B}  \frac{\dot R^2}{R^2} \nonumber\\ 
&&-\frac{k(n-1)}{R^2(1-B)(1+B)^{(n+3)/(n-1)}}
\ea
\n and
\be
\frac{16 \pi \rho}{n(n+1)} = \frac{\dot R^2}{R^2} + \frac{k}{R^2 
(1+B)^{\frac{n+3}{n-1}}}. 
\ee
\n The metric (17) asymptotically approaches the isotropic cosmological form 
in higher dimensions. The fluid flow is shear-free and irrotational. The
metric has a singularity at the origin. It represents an inhomogeneous
centro-symmetric cosmological model with
\be
\Theta = (n+1) \frac{\dot R}{R}, ~\dot u_r = - \frac{2(n-1)}{r(1 + k 
\frac{r^2}{4})} \frac{B}{1-B^2} 
\ee
\n as respectively kinematical parameters of expansion and acceleration.
The expressions (18) - (20) show the dimensional dependence of the 
physical and kinematical parameters.

\n In the metric (17) the parameter $B$ refers to the mass particle 
and its evolution in time is linked to the cosmological evolution through 
its dependence on the scale factor in eqn. (16). For $n>1$, which is 
always the case, it will go as inverse power of the scale factor. For the 
expanding phase it would decrease and the opposite would be the case 
for contracting phase. It also includes the following particular cases: 
(i) when $k=0, R(t) = const.$, it degenerates into the $(n+2)$
dimensional version of the Schwarzschild exterior metric, (ii) when $n=2$, 
it reduces to the McVittiee solution, (iii) when $m=0$, it represents 
$(n+2)$ dimensional Robertson-Walker universe and (iv) when $k = 1, R(t) 
= const.$, it describes mass particle placed in the $n+2$ dimensional 
static Einstein universe. \\

\n{\bf Acknowledgements:}  LKP and RT express their thanks to IUCAA
for facilities provided.

\end{document}